\documentclass[prl, showpacs, showkeys, twocolumn]{revtex4}
\usepackage{bm}
\usepackage{amssymb,amsmath,amsthm}
\usepackage{mathrsfs}
\begin{document}
\title{Explicit form of the Mann--Marolf surface term in (3+1) dimensions}
\author{Matt Visser}
\email{matt.visser@mcs.vuw.ac.nz}
\affiliation{School of Mathematics, Statistics, and Computer Science, 
Victoria University of Wellington, PO Box 600, Wellington, New Zealand\\}

\date{28 October 2008; 8 January 2009;
\LaTeX-ed \today}
\begin{abstract}
The Mann--Marolf surface term is a specific candidate for the ``reference background term'' that is to be subtracted from the Gibbons--Hawking surface term in order make the  total gravitational action of asymptotically flat spacetimes finite. That is, the total gravitational action is taken to be:\\
\null \; 
(Einstein--Hilbert bulk term) + (Gibbons--Hawking surface term) 
-- (Mann--Marolf surface term). \; \\
As presented by Mann and Marolf, their surface term is specified \emph{implicitly} in terms of the Ricci tensor of the boundary.  Herein I demonstrate that for the physically interesting case of a (3+1) dimensional bulk spacetime, the Mann--Marolf surface term can be specified \emph{explicitly} in terms of the Einstein tensor of the (2+1) dimensional boundary.

\end{abstract}


\maketitle

\def\d{{\mathrm{d}}}
\def\implies{\Rightarrow}
\newcommand{\scri}{\mathscr{I}}
\newcommand{\sun}{\ensuremath{\odot}}
\def\tr{{\mathrm{tr}}}
\def\sech{{\mathrm{sech}}}
\def\etc{\emph{etc}}
\def\ie{{\emph{i.e.}}}
\def\eg{{\emph{e.g.}}}
\newtheorem{theorem}{Theorem}
\newtheorem{corollary}{Corollary}
\def\lint{\hbox{\Large $\displaystyle\int$}} 
\def\hint{\hbox{\Huge $\displaystyle\int$}}  
\def\Re{{\mathrm{Re}}}
\def\Im{{\mathrm{Im}}}
\def\R{{\mathcal{R}}}
\def\G{{\mathcal{G}}}
\section{Introduction}

It has been known for many years that the Einstein--Hilbert term is only part of the story when it comes to considering the gravitational action. For a bulk region $\Omega$ bounded by a surface $\partial\Omega$ the full gravitational action is of the form~\cite{gibbons-hawking}
\begin{eqnarray}
\label{E:total}
S &=& -{1\over16\pi} \int_\Omega R(g_4) \; \sqrt{-g_4} \; \d^4 x 
\nonumber
\\
&&
- {1\over8\pi}\oint_{\partial\Omega} \{ K  + B(g_3) \}\;  \sqrt{-g_3} \; \d^3 x.
\end{eqnarray}
The extrinsic curvature term $K$ is commonly called the Gibbons--Hawking term (or sometimes the Gibbons--Hawking--York term). This term, and its normalization relative to the bulk term, is designed to ensure that as long as the induced metric on the boundary is held fixed, variations in the action depend only on the change  in the bulk metric~\cite{gibbons-hawking}
\begin{equation}
\label{E:variation}
\delta S = -{1\over16\pi} \int_\Omega G^{ab}(g_4)\; [\delta g_4]_{ab} \; \sqrt{-g_4} \; \d^4 x. 
\end{equation}
In contrast, the remaining term $B(g_3)$ is a ``reference term'' (also called a counterterm)  that depends only on the intrinsic geometry of the boundary, and is used to set the zero for the gravitational action~\cite{gibbons-hawking}. That is, the $B(g_3)$ counterterm is used to make the total action finite when evaluated on classical solutions --- these are configurations which we expect to contribute significantly in the stationary phase approximation to the path integral, so it would be desirable to ensure that they are assigned a finite phase~\cite{gibbons-hawking}. In particular, one would like the total gravitational action of Minkowski spacetime to be zero --- and this is enough to force you to realise that \emph{some} nonzero prescription for $B(g_3)$ is essential.

Over the years many suggestions have been made for the form of this ``reference term'' $B(g_3)$. For instance: If the boundary $\partial \Omega$, with its induced metric $g_3$, can be isometrically embedded into flat Minkowski spacetime, then Gibbons and Hawking argued that it is most useful to choose
\begin{equation}
\label{E:gh}
B(g_3) = - \hat K,
\end{equation}
where $\hat K$ is the extrinsic curvature of the boundary when it is isometrically embedded in flat Minkowski spacetime~\cite{gibbons-hawking}. For technical reasons the boundary is first placed at ``finite distance'', and the counter-term evaluated there.  Only then is a limiting procedure adopted to take the boundary to asymptotic infinity~\cite{gibbons-hawking}. 

Unfortunately there are many interesting situations (\eg, the Kerr spacetime) where any boundary placed at ``finite distance'' has a complicated intrinsic geometry that \emph{cannot} be isometrically embedded into flat Minkowski spacetime, and in these situations another reference prescription is called for.  A particularly promising recent suggestion is the Mann--Marolf boundary term introduced in~\cite{mann-marolf}, and further discussed in~\cite{mann-marolf2, mann-marolf3, mann-et-al, abhay, others}.

The Mann--Marolf counterterm is in many ways an example of a Zen k\"oan --- ``subtract the counterterm that is no counterterm''.  Specifically, define a ``virtual extrinsic curvature'', and subtract the trace of this ``virtual extrinsic curvature''. Adopting the definition
\begin{equation}
\label{E:mm}
\R_{ij}(g_3) = \hat K_{ij} \; \hat K - \hat K_i{}^m \; \hat K_{mj},
\end{equation}
the ``virtual extrinsic curvature'' $\hat K_{ij}$  is what the extrinsic curvature of  the boundary ``would have been'' if it were possible to embed the boundary into a flat Minkowski spacetime. This \emph{implicitly} defines $\hat K_{ij}$ as a function of the (2+1) Ricci tensor $\R_{ij}$ on the boundary, and thus \emph{implicitly} defines $\hat K_{ij}$ as a function of the induced (2+1) dimensional boundary geometry. We shall now seek to make this relation \emph{explicit}.

\section{Explicit evaluation of the Mann--Marolf counterterm} 

First raise one index on the defining relation
\begin{equation}
\label{E:mm2}
\R_{i}{}^{j} = \hat K_{i}{}^{j} \; \hat K - \hat K_i{}^m \; \hat K_{m}{}^{j}.
\end{equation}
Now pick some point on the boundary, and choose coordinates to diagonalize the tensor $\hat K_{i}{}^{j}$ at that point by a similarity transformation. This can always be done --- except possibly for a set of measure zero where the tensor takes on one of the non-trivial Jordan canonical forms. To deal with this exceptional case simply add a small perturbation to $\hat K_{i}{}^{j}$ to lift any degeneracy there may be between the eigenvalues of $\hat K_{i}{}^{j}$.  This guarantees that the perturbed tensor is diagonalizable by similarity transformations, and one can work with the perturbed diagonalized $\hat K_{i}{}^{j}$ up to the penultimate stage of the argument, and then set the perturbation to zero. Thus there is no loss of generality in taking $\hat K_{i}{}^{j}$ to be diagonalizable. But if $\hat K_{i}{}^{j}$ is diagonal, then automatically $\R_{i}{}^{j}$ is also diagonal.  

In Euclidean signature this appeal to the Jordan canonical form could have been short-circuited by first going to an orthonormal basis for the 3-metric, thereby setting $g_{\hat i\hat j} = \delta_{\hat i\hat j}$, and then using orthogonal transformations to diagonalize the symmetric matrix $\hat K_{\hat i \hat j}$. However this procedure fails in Lorentzian signature --- one has to classify all possible canonical forms for $3\times3$ symmetric matrices under local Lorentz transformations, and their stability under eigenvalue perturbations. In Lorentzian signature it is more direct to consider the non-symmetric mixed tensor $\hat K_{i}{}^{j}$  and appeal to the Jordan canonical form argument as above.

Up to this stage we could have worked in any number of dimensions $d\geq 4$, wheras  $d=4$, corresponding to (3+1) dimensions in the bulk and (2+1) dimensions in the boundary, is physically the most important case. Mathematically this is also the most difficult case to deal with, as $d=4$ does not exhibit the many technical simplifications that occur for $d>4$~\cite{mann-marolf, mann-marolf2, mann-marolf3, mann-et-al, abhay}.  However it is only in this physically most relevant situation of $d=4$ that we will succeed in inverting $\R(\hat K)$ to obtain $\hat K(\R)$.

When the bulk is (3+1) dimensional, so that the boundary is (2+1) dimensional, we can write the defining relations for $\hat K_{i}{}^{j}$  as:
\begin{eqnarray}
\label{E:mm3}
\R_{1}{}^{1} &=& \hat K_{1}{}^{1} \; \{\hat K_2{}^2 +  \hat K_{3}{}^{3} \};
\\
\R_{2}{}^{2} &=& \hat K_{2}{}^{2} \; \{\hat K_3{}^3 +  \hat K_{1}{}^{1} \};
\\
\R_{3}{}^{3} &=& \hat K_{3}{}^{3} \; \{\hat K_1{}^1 +  \hat K_{2}{}^{2} \}.
\end{eqnarray}
The Ricci scalar then satisfies
\begin{eqnarray}
\R &=& \R_{1}{}^{1} + \R_{2}{}^{2} + \R_{3}{}^{3} 
\\
&=& 
2\{  \hat K_{1}{}^{1} \;\hat K_2{}^2 + \hat K_{2}{}^{2} \;\hat K_3{}^3 
+ \hat K_{3}{}^{3} \;\hat K_1{}^1 \},
\end{eqnarray}
so for the Einstein tensor we have the particularly simple expressions
\begin{eqnarray}
\label{E:G1}
\G_{1}{}^{1} &=& -\hat K_2{}^2  \; \hat K_{3}{}^{3};
\\
\label{E:G2}
\G_{2}{}^{2} &=& -\hat K_3{}^3  \; \hat K_{1}{}^{1};
\\
\label{E:G3}
\G_{3}{}^{3} &=& -\hat K_1{}^1  \; \hat K_{2}{}^{2} .
\end{eqnarray}
Therefore
\begin{equation}
\G_{1}{}^{1} \; \G_{2}{}^{2} \; \G_{3}{}^{3}  = - 
\left\{ \hat K_{1}{}^{1} \; \hat K_{2}{}^{2} \; \hat K_{3}{}^{3} \right\}^2.
\end{equation}
These nonlinear algebraic relations can now be explicitly inverted to yield:
\begin{eqnarray}
\hat K_{1}{}^{1} &=& 
\sqrt{ 
- {\G_{2}{}^{2} \; \G_{3}{}^{3}
\over
\G_{1}{}^{1}}};
\\
\hat K_{2}{}^{2} &=&
 \sqrt{ - {
\G_{3}{}^{3}\; \G_{1}{}^{1}
\over
\G_{2}{}^{2}  }};
\\
\hat K_{3}{}^{3} &=&
 \sqrt{ - {
\G_{1}{}^{1} \;\G_{2}{}^{2}
\over
\G_{3}{}^{3}  }}.
\end{eqnarray}
Note that the sign of the square root has been chosen to ultimately be compatible with the Gibbons--Hawking prescription when the boundary is isometrically embeddable in Minkowski spacetime.  Note further that there is a risk of a ``divide by zero'' if one of the denominators happens to vanish. (We shall soon see that this apparently technical point actually arises in surprisingly simple situations --- such as static spherically symmetric spacetimes. Multiplying both the numerator and denominator by appropriate factors we see
\begin{eqnarray}
\hat K_{1}{}^{1} 
&=&
{\sqrt{ - \G_{1}{}^{1}  \; \G_{2}{}^{2} \; \G_{3}{}^{3} } 
\over \G_{1}{}^{1} }
 = 
{\G_{2}{}^{2}  \; \G_{3}{}^{3} 
\over \sqrt{ - \G_{1}{}^{1}  \; \G_{2}{}^{2} \; \G_{3}{}^{3} } }
 ;
\\
\hat K_{2}{}^{2} &=&
{\sqrt{ - \G_{1}{}^{1}  \; \G_{2}{}^{2} \; \G_{3}{}^{3} } 
\over \G_{2}{}^{2} }
 = 
{\G_{3}{}^{3}  \; \G_{1}{}^{1} 
\over \sqrt{ - \G_{1}{}^{1}  \; \G_{2}{}^{2} \; \G_{3}{}^{3} } }
 ;
\\
\hat K_{3}{}^{3} &=&
{\sqrt{ - \G_{1}{}^{1}  \; \G_{2}{}^{2} \; \G_{3}{}^{3} } 
\over \G_{3}{}^{3} }
 = 
{\G_{1}{}^{1}  \; \G_{2}{}^{2} 
\over \sqrt{ - \G_{1}{}^{1}  \; \G_{2}{}^{2} \; \G_{3}{}^{3} } }.
\end{eqnarray}
But here we recognize  the determinant of the (2+1) Einstein tensor. That is, (using $\bullet$ symbols for index placeholders),
\begin{eqnarray}
\hat K_{1}{}^{1} &=& 
{\sqrt{ \det(-\G_{\bullet}{}^{\bullet})}\over\G_{1}{}^{1}}
= 
{\G_{2}{}^{2}  \; \G_{3}{}^{3} \over \sqrt{ \det(-\G_{\bullet}{}^{\bullet})}}
; 
\\
\hat K_{2}{}^{2} &=&
{\sqrt{ \det(-\G_{\bullet}{}^{\bullet})}\over \G_{2}{}^{2}}
=
{\G_{3}{}^{3}  \; \G_{1}{}^{1} \over \sqrt{ \det(-\G_{\bullet}{}^{\bullet})}}
;
\\
\hat K_{3}{}^{3} &=&
{\sqrt{ \det(-\G_{\bullet}{}^{\bullet})}\over\G_{3}{}^{3}}
=
{\G_{1}{}^{1}  \; \G_{2}{}^{2} \over \sqrt{ \det(-\G_{\bullet}{}^{\bullet})}}.
\end{eqnarray}
Finally, unwrapping the diagonalization, we have 
\begin{equation}
\label{E:vv}
\hat K_i{}^j = \sqrt{ \det(-\G_{\bullet}{}^{\bullet})} \; [ (\G_{\bullet}{}^{\bullet})^{-1}]_i{}^j,
\end{equation}
and
\begin{equation}
\label{E:vb1}
\hat K = \sqrt{ \det(-\G_{\bullet}{}^{\bullet})} \; \tr[ (\G_{\bullet}{}^{\bullet})^{-1}].
\end{equation}
This is the first version of the key result --- it \emph{explicitly} yields the Mann--Marolf surface terms in terms of the determinant and trace of the inverse of the (2+1) Einstein tensor on the boundary of the spacetime. The ``divide by zero'' issues we alluded to earlier are now seen to be related to the possible vanishing of $\det(\G_{\bullet}{}^{\bullet})$. 

Alternatively, if one wishes to sidestep the explicit matrix inversion, a second (equivalent) version of our key result is to use a combination of Cramer's rule and the Laplace expansion for a determinant to derive
\begin{equation}
\label{E:vv2}
\hat K_i{}^j = { \varepsilon_{imn} \;  \G_p{}^m \; \G_q{}^n\;  \varepsilon^{pqj} \over
\sqrt{ \det(-\G_{\bullet}{}^{\bullet})} },
\end{equation}
and
\begin{equation}
\label{E:vb2}
\hat K =  {\G^2 -  \G_{i}{}^{j} \; \G_{j}{}^{i} 
\over 2\sqrt{ \det(-\G_{\bullet}{}^{\bullet})}},
\end{equation}
where $\G = -{1\over2} \R$ is the trace of the Einstein tensor. 
This can also be recast directly in terms of the Ricci tensor in the form
\begin{equation}
\label{E:vv3}
\hat K_i{}^j = { \varepsilon_{imn} \;  \R_p{}^m \; \R_q{}^n\;  \varepsilon^{pqj} 
- {1\over2} \R \; \R_i{}^j  - {1\over 4} \R^2 \; \delta_i{}^j 
\over
\sqrt{ \det(-\R_{\bullet}{}^{\bullet}+{1\over2}\, \R \, \mathbf{I})} },
\end{equation}
and
\begin{equation}
\label{E:vb3}
\hat K =  {\R^2 -  2\R_{i}{}^{j} \; \R_{j}{}^{i} 
\over 4\sqrt{ \det(-\R_{\bullet}{}^{\bullet}+{1\over2} \,\R \;\mathbf{I})}},
\end{equation}
this providing the third version of our key result. 

A fourth version of our key result can be derived by considering the quantity $\det(-\G_{\bullet}{}^{\bullet}+\epsilon\; \mathbf{I}_{\bullet}{}^{\bullet})$ and then noting
\begin{equation}
\label{E:vb4}
\hat K =  {2} \; {\d\over\d\epsilon}\left.  
\sqrt{ \det(-\G_{\bullet}{}^{\bullet}+\epsilon\; \mathbf{I}_{\bullet}{}^{\bullet}) }
\right|_{\epsilon=0}.
\end{equation}

\paragraph{Exceptional case 1:}
If in these formulae one encounters a ``divide by zero'' error it can only be because one (or more) of the eigenvalues of the Einstein tensor  $\G_\bullet{}^\bullet$ is zero. But then, by Eqs.~(\ref{E:G1})--(\ref{E:G3}) at least two of the eigenvalues are zero. If the Einstein tensor is singular but not zero, we can without loss of generality  take the non-zero eigenvalue to be $\G_1{}^1$ in which case 
\begin{equation}
\G_{1}{}^{1} = -{1\over2}\R, \qquad \G_{2}{}^{2} = \G_{3}{}^{3} =  0.
\end{equation}
But then we must have $\hat K_1{}^1=0$, and the only nontrivial constraint comes from Eq.~(\ref{E:G1}), which then implies
\begin{equation}
\label{E:special}
\hat K_{1}{}^{1} = 0 , \qquad 
\hat K_{2}{}^{2} = \sqrt{\R\over2} e^\vartheta, \qquad
\hat K _{3}{}^{3} =   \sqrt{\R\over2} e^{-\vartheta},
\end{equation}
whence
\begin{equation}
\label{E:special2}
\hat K = \sqrt{  2 \R } \; \cosh\vartheta.
\end{equation}
Only in situations of additional symmetry might we be able to \emph{guarantee} more about the parameter $\vartheta$. However, noting that the whole philosophy behind the Mann--Marolf counterterm is simply to write down \emph{a} solution for the virtual extrinsic curvature defined by Eq.~(\ref{E:mm}), in the current situation no-one can stop us from simply \emph{choosing} $\vartheta=0$, in which case 
\begin{equation}
\label{E:special3}
\hat K_{1}{}^{1} = 0 , \qquad 
\hat K_{2}{}^{2} = \hat K _{3}{}^{3}  = \sqrt{\R\over2} , \qquad 
\hat K = \sqrt{  2 \R }.
\end{equation}
It is then easy to check that this is compatible with the appropriate limiting case of Eq. (\ref{E:vb4}).

\paragraph{Exceptional case 2:}
Finally in the case that the Einstein tensor is zero, (which due to the special features of 3 dimensions implies in particular that the boundary is Riemann flat), Eqs.~(\ref{E:G1})--(\ref{E:G3}) imply that at least two of the eigenvalues of $\hat K_\bullet{}^\bullet$ are zero.  Only in situations of additional symmetry might we be able to \emph{guarantee} more about the one remaining nonzero eigenvalue of $\hat K_\bullet{}^\bullet$. However, as before, noting that the whole philosophy behind the Mann--Marolf counterterm is simply to write down \emph{a} solution for the virtual extrinsic curvature defined by Eq.~(\ref{E:mm}), in the current situation no-one can stop us from simply \emph{choosing} $\hat K_\bullet{}^\bullet=0$ and  $\hat K=0$. Since in this sub-case the Ricci tensor is also zero, it is again easy to see that this is compatible with the appropriate limiting case of Eq. (\ref{E:vb4}).

\paragraph{In short:} If one wishes to calculate the entire ``virtual extrinsic curvature'' one must deal with Eq.~(\ref{E:vv}), (\ref{E:vv2}), or (\ref{E:vv3}). If one is satisfied with just knowing the boundary term itself, the trace of the ``virtual extrinsic cuvature'', then any of the formulae (\ref{E:vb1}), (\ref{E:vb2}),  (\ref{E:vb3}),  or (\ref{E:vb4}) suffice.  At the very worst one might have to deal with the special case of Eq.~(\ref{E:special3}), which is compatible with the appropriate limit of  Eq. (\ref{E:vb4}).

\section{Comparison with other surface terms} 

The Mann--Marolf counterterm has already been extensively compared with other counterterms appearing in the literature~\cite{mann-marolf, mann-marolf2, mann-marolf3, mann-et-al}, so at this stage the only point of consistency checking is to verify that our explicit formulae make sense and yield the expected results.

\noindent i) 
Simply from the way it is defined, it is clear that if the boundary is isometrically embeddable in Minkowski spacetime, then the Mann--Marolf procedure automatically reproduces the Gibbons--Hawking prescription~\cite{gibbons-hawking}.

\noindent ii) 
To be more explicit about this: Consider the special case of a static spherically symmetric geometry, adopt Schwarzschild curvature coordinates, and place the boundary surface at $r=r_*$. One then has
\begin{equation}
\R_{1}{}^{1} = 0; \qquad \R_{2}{}^{2} =  \R_{3}{}^{3} = {1\over r_*^2}. 
\end{equation}
So for the boundary Einstein tensor
\begin{equation}
\G_{1}{}^{1} = -{1\over r_*^2}; \qquad \G_{2}{}^{2} =  \G_{3}{}^{3} = 0. 
\end{equation}
This implies that  this is one of those situations where $\det(\G_{\bullet}{}^{\bullet})$ vanishes. Using spherical symmetry, $\vartheta=0$ and Eq.~(\ref{E:special}) quickly yields
\begin{equation}
\hat K = {2\over r_*},
\end{equation}
which is compatible with the Gibbons--Hawking calculation in~\cite{gibbons-hawking}.

\noindent iii)  
Next consider an arbitrary  static spacetime --- and pick the boundary to be one of the level-surfaces of  $g_{tt}= -N^2$.  Then one has $g_3 = -N^2 \oplus g_2$ and 
\begin{equation}
\R_{1}{}^{1} = 0; \qquad  \R_{2}{}^{2} =  \R_{3}{}^{3} = {1\over 2} \R. 
\end{equation}
Then for the boundary Einstein tensor
\begin{equation}
\G_{1}{}^{1} = -{1\over2}\R, \qquad \G_{2}{}^{2} = \G_{3}{}^{3} =  0.
\end{equation}
This  is another of those situations where $\det(\G_{\bullet}{}^{\bullet})$ vanishes.  Eq.~(\ref{E:special3})  now yields
\begin{equation}
\hat K = \sqrt{2 R},
\end{equation}
which is compatible with the original Lau-Mann counterterm investigated in~\cite{lau, mann-alone}.

\noindent iv)  Finally, another popular counterterm is the Kraus--Larsen--Siebelink counterterm~\cite{kraus-larsen-siebelink}
\begin{equation}
B_\mathrm{KLS} = -{\R^{3/2}\over \sqrt{ \R^2 - \R_i{}^j \, \R_j{}^i }},
\end{equation}
which in diagonalized form is
\begin{equation}
B_\mathrm{KLS} = -{ ( \R_{1}{}^{1} + \R_{2}{}^{2} + \R_{3}{}^{3} ) ^{3/2}
\over
 \sqrt{   2 \{ \R_{1}{}^{1}  \; \R_{2}{}^{2} + \R_{2}{}^{2}  \; \R_{3}{}^{3} +
\R_{3}{}^{3}  \; \R_{1}{}^{1} \} }}.
\end{equation}
While this is clearly closely related to the Mann--Marolf counterterm, it is also clear that it will not \emph{equal} the Mann--Marolf counterterm except in cases of exceptionally high symmetry. For example, for static spacetimes as considered above,  
\begin{equation}
B_\mathrm{KLS} \to -\sqrt{2 R}=-\hat K.
\end{equation}
Thus the KLS, Lau-Mann, and Mann--Marolf counterterms are in this situation all identical.

\section{Implications} 

In assessing various proposed surface counterterms it is very useful to have explicit formulae available, since only with explicit formulae at hand may it be easy to fully understand the possible benefits and pitfalls of the various prescriptions. The Mann--Marolf counterterm is subject to two significant pitfalls that could not easily be determined in the absence of the explicit formulae derived above:
\begin{itemize}
\item If $\det(\G_{\bullet}{}^{\bullet})$ vanishes, then one has to adopt a special case analysis,  whose genesis would not be at all obvious if all one has on hand is the basic defining relation of equations (\ref{E:mm}) or (\ref{E:mm2}).
\item  If $\det(\G_{\bullet}{}^{\bullet})$ is positive then the ``virtual extrinsic curvature'' is unavoidably complex. This is a somewhat unexpected feature that would be difficult to interpret, or even detect,  if all one has on hand is the basic defining relation of equations (\ref{E:mm}) or (\ref{E:mm2}).
\end{itemize}
An interesting side-effect of the above is that any 3 geometry that is isometrically embeddable in flat 4 space must satisfy $\det(\G_{\bullet}{}^{\bullet})\leq 0$. 

Finally, it should be pointed out that the results of this note very definitely make use of the special properties of  3 geometries embedded in 4 geometries; there is no reason to expect that similar explicit formulae for the ``virtual extrinsic curvature'' $\hat K_{ij}$ will persist for $d>4$, and any attempt at performing an analysis along the lines presented above quickly dies in a morass of intractable nonlinear algebraic equations. Fortunately the $d=4$ case, [that is, the (3+1)-dimensional case], is physically the most interesting one.

\section*{Acknowledgments}
This research was supported by the Marsden Fund administered by the Royal Society of New Zealand. 



\end{document}